\documentstyle[preprint,aps,eqsecnum]{revtex}
\begin{document}
\draft
%\preprint{dvi file made on \today}
\title
{Transport Properties of Solitons}
\author{A.~H.~Castro Neto}
\bigskip
%
%\begin{instit}
\address
{Department of Physics,
University of Illinois at Urbana-Champaign\\
1100 W.Green St., Urbana, IL, 61801-3080, U.S.A.}
%\end{instit}
%
\author{A.~O.~Caldeira}
\bigskip
\address{
Instituto de F\'isica ``Gleb Wataghin"\\
Departamento de F\'isica do Estado S\'olido e Ci\^encia dos Materiais\\
Universidade Estadual de Campinas\\
13081, Campinas, SP, Brazil\\}

\maketitle

\bigskip

\begin{abstract}
We calculate in this article the transport coefficients which characterize
the dynamics of solitons in quantum field theory using the methods
of dissipative quantum systems. We show how the damping and
diffusion coefficients of soliton-like excitations can be calculated
using the integral functional formalism. The model obtained in
this article has new features which cannot be obtained in the
standard models of dissipation in quantum mechanics.
\end{abstract}

\bigskip


\narrowtext

\section{Introduction}

There are still controversies about the calculation of transport properties
involving soliton-like particles (see \cite{david} and references therein).
It is not
completely clear how the thermalization of collective excitations, such
as solitons, can be carried out. In
general, the pathway is based on heuristic approaches which take into account
averages over configurations without going deeply inside the dynamical
characteristics of the formation of a collective excitation.

The main goal of this article is to link two fields of study which seem to
be far apart up to now but, as we shall show, are really closely related.

One of these areas is the quantum theory of solitons developed in the
seventies \cite{raja} which teaches us how to quantize classical theories
which have solitons as solutions of their equations of motion. The other
area is the field of dissipative quantum systems which had a great
development during the eighties \cite{amir}.

When we study the dynamics of a classical particle moving through a viscous
environment we learn that one can describe its motion using two parameters,
the damping and the diffusion coefficient. The former is related to the
systematic force applied on the particle by the environment and the latter
is related with the fluctuations due to the interaction (for a clear discussion
about this issue see \cite{reif}).

It was shown some years ago that the same physical situation occurs in quantum
systems \cite{amir}, therefore if a particle interacts with an environment such
as a set of decoupled harmonic oscillators (the oscillator model) its
motion is damped, which means in quantum mechanical terms that the center
of motion of its wavepacket undergoes a damped motion while its width increases
with time. Exactly as in the classical problem all the transport properties
of the quantum system can be described by a damping and a diffusion
coefficient.

One of the main results of the standard approach is that the damping time is
independent
of temperature and even at zero temperature the motion of the particle is
damped. As will show in this article this behaviour is not expect for solitons
since the soliton and the environment have the same microscopic origin.

{}From the point of view of quantum field theory, classical and translational
invariant theories with solitons can be quantized by the collective
coordinate method \cite{raja}. In the classical theory, solitons are field
configurations which move in space without changing its shape. Therefore, the
knowledge of the position of the center of soliton as a function of time
is enough to describe its dynamics. In quantum field theory the center of
the soliton plays a special role. The center of the soliton can be viewed as
a true quantum dynamical variable and we describe the soliton as a particle.
The problem here is that at finite temperatures not all the degrees of freedom
(infinite, indeed) collaborated in the formation of the soliton and
therefore
this soliton is never free to move as in the classical theory, that is,
there is always a residual interaction due to quantum degrees of freedom
which changes
drastically the dynamical properties of the system. Our task is to show that,
under some approximations, this residual interaction gives rise to a damped
motion of the soliton, in other words, the soliton undergoes a brownian motion.

Form this point of view we are presenting here a new model for dissipation
in quantum mechanics and, as we would expect, very different from the
oscillator model described before.

We could have started from the definition of the model \cite{prl}, which
by itself is completely new in the field of quantum dissipation,
without mentioning
its origin in quantum field theory, but we think it is of interest to
show the generality of the model in this context.

We will show here how we can calculate the transport properties in this
framework, namely, the damping and the diffusion coefficient. We will
show that the damping time depends on the temperature, a result which
can not be obtained in the oscillator model.

In section II we present the results of the quantization of the solitons in
field theory and the approximations involved in our model. Although
the same approach can be found in great detail in Ref. \cite{raja}
we include this topic here for the sake of completeness.
In Section III we develop the
the integral functional formalism which allows us to calculate the
transport properties. Section IV contains our conclusions.

\section{Quantization of Solitons}

Let us consider a classical action in 1+1 dimensions for a scalar field,
\begin{equation}
S=\int d^2x \left(\frac{1}{2}\left(\partial^{\mu}\phi\right)\left(\partial_{
\mu}\phi\right)-U(g,\phi)\right)
\end{equation}
where $U$ is a potential function of the field $\phi$ and the coupling
constant $g$. Here we will assume that $U$ has a degenerate absolute
minimum. We can always rescale the field such as $\phi \to \phi/g$ and
the potential as $U \to U/g^2$ (by naive dimensional analysis) and
rewrite (2.1) as
\begin{equation}
S=\frac{1}{g^2} \int d^2x \left(\frac{1}{2} \left(\partial^{\mu}\phi\right)
\left(\partial_{\mu}\phi\right)-U(\phi)\right).
\end{equation}

Consider the static part of (2.2),
\begin{equation}
V=\frac{1}{g^2} \int d^2x \left(\frac{1}{2}\left(\frac{\partial \phi}
{\partial x}\right)^2+U(\phi)\right)
\end{equation}
and its extremum value given by $\delta V/\delta\phi_s=0$, namely,
\begin{equation}
\frac{d^2\phi_s}{dx^2}=\frac{dU(\phi_s)}{d\phi_s}
\end{equation}
which, due to the translational invariance of (2.1), must have the
form
\begin{equation}
\phi_s=\phi_s(x-x_0)
\end{equation}
where $x_0$ is a constant. Eq.(2.5) is a static localized solution of
the classical equations of motion.

We expand (2.3) around the extremum, (2.4), in terms of the coupling
constant $g$ as,
\begin{equation}
\phi(x,t)=\phi_s(x-x_0)+g \delta\phi(x,t)
\end{equation}
and one finds,
\begin{equation}
V[\delta\phi]=V[\phi_s]+\int d^2x \, \delta\phi(x,t) \left( -\frac{d^2}{dx^2}
+ \left(\frac{d^2U}{d\phi^2}\right)_{\phi_s}\right)\delta\phi(x,t)
\end{equation}
plus higher order terms in $g$.

Using (2.4) we get,
\begin{equation}
M=V[\phi_s]=\frac{1}{g^2} \int d^2x \left(\frac{d\phi_s}{dx}\right)^2
\end{equation}
is the classical soliton mass.

Therefore, up to leading order in $g$, the eigenmodes of the system are
given by
\begin{equation}
\left(-\frac{d^2}{dx^2}+\left(\frac{d^2U}{d\phi^2}\right)_{\phi_s}
\right)\Psi_n(x-x_0)=\Omega^2_n \Psi_n(x-x_0).
\end{equation}

Observe that the function $d\phi_s/dx$ is an eigenfunction with eigenvalue
zero (use (2.4) in (2.9)) indicating that we should expect divergencies in high
order terms in the perturbative expansion. The way to deal with this ``zero"
mode is the following: suppose we displace the center of the soliton, $x_0$,
by an infinitesimal amount $\delta x_0$. Thus, up to first
order in $\delta x_0$,
\begin{equation}
\delta_0\phi_s=\phi_s(x_0+\delta x_0)-\phi_s(x_0)=\frac{\partial\phi_s}
{\partial x_0} \delta x_0
\end{equation}
but since $\phi_s$ is a function of the difference $x-x_0$ we can rewrite
(2.10) as
\begin{equation}
\delta_0\phi_s=-\frac{d\phi_s}{dx} \delta x_0
\end{equation}
therefore the eigenmode $d\phi_s/dx$ is related with the movement of the
localized solution, that is, the motion of the soliton. Call,
\begin{equation}
\Psi_0(x-x_0)=\frac{1}{N} \frac{d\phi_s}{dx}
\end{equation}
where $N$ is a normalization constant.

We rewrite $\delta\phi$ in (2.6) as,
\begin{equation}
\delta\phi(x,t)=\sum_{n=0}^{\infty} a_n(t) \Psi_n(x-x_0).
\end{equation}
Therefore, from (2.6),
\begin{equation}
\phi(x,t)=\phi_s(x-x_0)+\frac{g}{N} a_0(t) \frac{d\phi_s}{dx}
+g \sum_{n=1}^{\infty} a_n(t) \Psi_n(x-x_0).
\end{equation}

Notice that the first two terms in r.h.s. of (2.14) appear in
the expansion obtained in (2.10). In
this way we see that the center of the soliton is a true dynamical variable
and we rewrite the expansion (2.14) as,
\begin{equation}
\phi(x,t)=\phi_s(x-x_0(t))+g \sum_{n=1}^{\infty} a_n(t) \Psi(x-x_0(t))
\end{equation}
which is known as the collective coordinate method. All the physics of this
system is present in (2.15). The soliton, described by the motion of its
center, is a collective excitation (since is a solution of the field
equation) but it is coupled to all other modes by the relative coordinate
$x-x_0$.

We can now rewrite the action (2.2) in the presence of the soliton using (2.14)
(see \cite{raja} for details) and obtain the classical Hamiltonian for
his problem,
\begin{equation}
H=-M +\frac{1}{2M}\left(P-\sum_{n,m=1}^{\infty} G_{mn} p_n q_m \right)^2
+\sum_{n=1}^{\infty}\left(\frac{p_n^2}{2}+\frac{\Omega_n^2 q^2_n}{2} \right)
\end{equation}
plus higher order terms in $g$. $P$ is the momentum canonically conjugated
to $x_0$, $q_n$ and $p_n$ are the conjugated pairs for the modes and
\begin{equation}
G_{mn}=-\int dx \, \frac{d\Psi_m}{dx}(x) \Psi_n(x)
\end{equation}
couples the soliton to the other modes (mesons) in the system.

It is clear now that the soliton can not move freely, the mesons are
scattered by the soliton which will move as a brownian particle. Although
(2.16) is perturbative in the coupling constant, we will assume that the main
physics of the problem is present in that expression.

The quantization of (2.16) is straightforward. We impose the commutation
relations
\begin{equation}
[\hat{x}_0,\hat{P}]=i\hbar
\end{equation}
\begin{equation}
[\hat{q}_n,\hat{p}_m]=i \hbar \delta_{nm}
\end{equation}
and write (2.16) in the symmetrized form (apart from the constant mass
term)
\begin{equation}
H= \frac{1}{2M}\left(\hat{P}-\hat{P}_{me}\right)^2 + \sum_{n=1}^{\infty}
\hbar \Omega_n \hat{b}^{\dag}_n \hat{b}_n
\end{equation}
where
\begin{equation}
\hat{b}_n=\left(\frac{\Omega_n}{2 \hbar}\right) \left(\hat{q}_n+i
\frac{\hat{p}_n}
{\Omega_n}\right)
\end{equation}
which obeys the usual commutation relations for bosons,
\begin{equation}
\left[\hat{b}_n,\hat{b}^{\dag}_m\right]=\delta_{nm} \, \, \, ;
\left[\hat{b}_n,\hat{b}_m\right]=0.
\end{equation}

In (2.20), $\hat{P}_{me}$ is the momentum of the mesonic field which is given
by
\begin{equation}
\hat{P}_{me}=\hat{P}_d+\hat{P}_{nd}
\end{equation}
where
\begin{equation}
\hat{P}_d=\sum_{n,m=1}^{\infty} \frac{\hbar}{2i} \left(\left(\frac{\Omega_n}
{\Omega_m}
\right)^{1/2}+\left(\frac{\Omega_m}{\Omega_n}\right)^{1/2}\right) G_{mn}
\hat{b}^{\dag}_m\hat{b}_n
\end{equation}
and
\begin{equation}
\hat{P}_{nd}=\sum_{n,m=1}^{\infty} \frac{\hbar}{4i} \left(\left(\frac{\Omega_n}
{\Omega_m}
\right)^{1/2}-\left(\frac{\Omega_m}{\Omega_n}\right)^{1/2}\right)G_{mn} \left(
\hat{b}_m\hat{b}_n-\hat{b}^{\dag}_m\hat{b}^{\dag}_n\right).
\end{equation}

The Hamiltonian (2.20) is well known in works on polaron dynamics
\cite{holstein,prb} and many perturbative approaches has been used in
order to understand its physical contents.

Observe that the $P_{nd}$ does not commute with the total number of
mesons
\begin{equation}
\hat{N}=\sum_{n=1}^{\infty}\hat{b}^{\dag}_n\hat{b}_n
\end{equation}
and therefore it breaks the number conservation. Actually this term is
related with high frequency oscillations in which we are not interested.
Our task is to show that in the sector of the Hilbert space where the
number $\hat{N}$ is conserved the physics of (2.20) is that of a brownian
particle. Therefore our new model in the context of quantum dissipation
is based on the following Hamiltonian \cite{prl}
\begin{equation}
H=\frac{1}{2M} :\left(\hat{P}-\sum_{n,m=1}^{\infty} \hbar g_{mn}
\hat{b}^{\dag}_n \hat{b}_n\right)^2:
+\sum_{n=1}^{\infty} \hbar \Omega_n \hat{b}^{\dag}_n \hat{b}_n
\end{equation}
where $:...:$ means normal order and
\begin{equation}
g_{mn}=\frac{1}{2i} \left(\left(\frac{\Omega_n}{\Omega_m}\right)^{1/2}
+\left(\frac{\Omega_m}{\Omega_n}\right)^{1/2}\right) G_{mn}
\end{equation}
is the coupling constant of the theory.

We believe that Hamiltonian (2.27) describes the main physics of the
dynamics of solitons at low energies and, by itself, can be used to
model other physical systems where the coupling of the particle and
the environment can not be described by the standard model.

\setcounter{equation}{0}
\section{Functional Integral Method}

\indent
The starting point for the calculations of the transport properties
of the soliton is the well-known Feynman-Vernon formalism \cite{feynman}
that the authors have recently applied \cite{prl} to the Hamiltonian
(2.27).

\indent
We are interested only in the quantum statistical properties of the soliton
and the mesons act only as a source of relaxation and diffusion
processes. Consider the density operator for the system soliton plus
mesons, $\hat{\rho}(t)$. This operator evolves in time according to

\begin{equation}
\hat{\rho}(t) = e^{-i\hat{H}t/\hbar} \hat{\rho}(0) e^{i\hat{H}t/\hbar}
\end{equation}

\noindent
where $\hat{H}$ is given by (2.27) and $\hat{\rho}(0)$ is the density
operator at $t = 0$ which we will assume to be decoupled as a product
of the soliton density operator, $\hat{\rho}_S(0)$, and the meson
density operator, $\hat{\rho}_R(0)$,

\begin{equation}
\hat{\rho}(0) = \hat{\rho}_S(0) \hat{\rho}_R(0)
\end{equation}

\noindent
where the symbol $\underline{S}$ refers to the soliton (system of
interest) and $\underline{R}$ to the mesons (the reservoir of excitations).

\indent
 We will consider that the mesons are in thermal equilibrium at $t=0$,
that is,
\begin{equation}
\hat{\rho}_R(0) = \frac{e^{-\beta \hat{H}_R}}{Z}
\end{equation}
where
\begin{equation}
Z = tr_R \left(e^{-\beta \hat{H}_R}\right)
\end{equation}
with
\begin{equation}
\beta = \frac{1}{K_BT}.
\end{equation}
Here $tr_R$ denotes the trace over the mesons variables and $K_B$ is
the Boltzmann constant. $\hat{H}_B$ is the free meson Hamiltonian
which is given by the last term on the right hand side of (2.27).

As we said, we are interested only in the quantum dynamics of the system S,
so, we define a reduced density operator
\begin{equation}
\hat{\rho}_s(t) = tr_R (\hat{\rho}(t))
\end{equation}
which contains all the information about $\underline{S}$
when it is in contact with $\underline{R}$.

Projecting now (3.1) in the coordinate representation of the soliton system
\cite{golds}
\begin{equation}
\hat{x}_o \mid q > = q \mid q >
\end{equation}
and in the coherent state representation for bosons (the mesons),
\begin{equation}
\hat{b}_n \mid \alpha_n > = \alpha_n \mid \alpha_n >
\end{equation}
we have \cite{castro neto} (see ref.\cite{prb} for details)
\begin{equation}
\rho_s (x, y, t) = \int{dx'} \int{dy'} J(x, y, t; x', y',0) \rho_s(x',y',0)
\end{equation}

Here we have used (3.2), (3.3), (3.6) and the completeness relation for the
representations above, namely,
\begin{equation}
\int{dq} \mid q > < q \mid = 1
\end{equation}
\begin{equation}
\int{\frac{d^2 \alpha}{\pi}} \mid \alpha > < \alpha \mid = 1
\end{equation}
where $d^2 \alpha = d(Re \alpha)d(Im \alpha)$ as usual.

In (3.9), $J$ is the superpropagator of the soliton, which can be written
as
\begin{equation}
J = \int^{x}_{x'}{Dx} \int^{y}_{y'}{Dy} \, \,
e^{\frac{i}{\hbar}(S_o[x]-S_o[y])} F[x,y]
\end{equation}
where
\begin{equation}
S_o[x] = \int^t_0{dt'} \; \left\{\frac{M_o \dot{x}^2(t')}{2}\right\}
\end{equation}
is the classical action for the free particle. F is the so-called
influence functional,
%\[
\begin{equation}
F[x,y] = \int{\frac{d^2 \vec{\alpha}}{\pi^N}} \int{\frac{d^2
\vec{\beta}}{\pi^N}} \int{\frac{d^2 \vec{\beta}'}{\pi^N}} \;
\rho_R(\vec{\beta}^*, \vec{\beta}') \; e^{-\mid \vec{\alpha} \mid^2 -
\frac{\mid \vec{\beta} \mid^2}{2} - \frac{\mid \vec{\beta}'\mid^2}{2}}
%\]
%\begin{equation}
\int^{\vec{\alpha}^*}_{\vec{\beta}} {D^2 \vec{\alpha}}
\int^{\vec{\alpha}}_{\vec{\beta}'^*} {D^2 \vec{\gamma}} \; e^{S_I[x,
\vec{\alpha}] + S^*_I[y,\vec{\gamma}]}
\end{equation}
where $\vec{\beta}$ denotes the vector $(\beta_1, \beta_2,
\beta_3,..., \beta_N)$ and $S_I$ is a complex action related to the
reservoir plus interaction,
\begin{equation}
S_I[x,\vec{\alpha}] = \int^t_0{dt'} \left\{ \frac{1}{2}
\left( \vec{\alpha} \cdot
\frac{d \vec{\alpha}^*}{dt'} - \vec{\alpha}^* \cdot
\frac{d\vec{\alpha}}{dt'}\right)
- \frac{i}{\hbar} (H_R - \dot{x} h_I) \right\}
\end{equation}
with
\begin{equation}
H_R = \sum^\infty_{n=1} \hbar \Omega_n \alpha^*_n \alpha_n
\end{equation}
\begin{equation}
h_I = \sum^\infty_{n,m=1} \hbar g_{nm} \alpha^*_m \alpha_n.
\end{equation}

In our case the lagrangean formalism simplifies the problem
transforming a nonlinear problem into a linear one. The action (3.15) is
quadratic in $\vec{\alpha}$, so it can be solved exactly. Observe
that the Euler-Lagrange equations for (3.15) are
\begin{equation}
\dot{\alpha}_n + i\Omega_n \alpha_n - i \dot{x} \sum^\infty_{m=1}
g_{mn}\alpha_m = 0
\end{equation}
\begin{equation}
\dot{\alpha}^*_n - i \Omega_n \alpha^*_n + i \dot{x}
\sum^\infty_{m=1} g_{nm} \alpha^*_m = 0
\end{equation}
which must be solved subject to the boundary conditions
\begin{equation}
\alpha_n(0) = \beta_n
\end{equation}
\begin{equation}
\alpha^*_n(t) = \alpha^*_n.
\end{equation}

Due to (2.17) we have $g_{nn} = 0$ and the modes are not coupled among
themselves. This makes (3.18) and (3.19) easy to solve. That set of
equations represents a set of harmonic oscillators forced by the
presence of the soliton. The result can be written as,
\begin{equation}
\alpha_n(\tau) = e^{-i \Omega_n \tau} \left(\beta_n + \sum^\infty_{m=1}
W_{nm}(\tau) \beta_m \right)
\end{equation}
\begin{equation}
\alpha^*_n(\tau) = e^{i \Omega_n\tau} \left(\alpha^*_n e^{-i \Omega_nt}
 +\sum^\infty_{m=1} \tilde{W}_{nm}(\tau) e^{-i \Omega_m t} \alpha^*_m \right)
\end{equation}
where $W_{nm}$ and $\tilde{W}_{nm}$ are functionals of $x(t)$
which obey the following equations
\begin{equation}
W_{nm}(\tau) = \delta_{nm} + \sum^\infty_{n'=1} \int_{0}^{\tau} dt' \,
K_{nn'}(t') W_{n'm}(t')
\end{equation}
\begin{equation}
\tilde{W}_{nm}(\tau) = \delta_{nm} + \sum^\infty_{n'=1} \int_{\tau}^t dt' \,
K_{nn'}(t') \tilde{W}_{n'm}(t')
\end{equation}
where
\begin{equation}
K_{nm}([x], \tau) = i g_{nm} \dot{x}(t') e^{i(\Omega_n - \Omega_m)t'}
\end{equation}
is the kernel of the integral equation
(observe that: $W_{nm}(t) = \tilde{W}_{mn}(0)$).

Now we expand the action (3.15) around the classical solution (3.22)
and (3.23) and obtain, after some integrations in (3.14)
\begin{equation}
F[x,y] = \prod^\infty_{n=1} (1-\Gamma_{nn}[x,y] \; \overline{n}_n)^{-1}
\end{equation}
where
\begin{equation}
\Gamma_{nm} = W^*_{nm}[y] + W_{mn}[x] + \sum^\infty_{\ell=1}
W^*_{\ell m}[y] W_{\ell n}[x]
\end{equation}
with
\begin{equation}
\overline{n}_n = \left(e^{\beta \hbar \Omega_n} - 1\right)^{-1}.
\end{equation}

Notice that (3.27) and (3.28) are exact, no approximations have been
made so far.

We see from (3.24) and (3.26) that $W_{nm}$ can be expressed as a power
series expansion of the Fourier transform of the soliton velocity,
$\dot{x}$, so for small soliton velocities we expect that only few terms
in (3.24) will be sufficient for a good description of the soliton dynamics.

Another way to see this is to notice that (3.24) and (3.25) are the
scattering amplitudes from the mode $k$ to the mode $j$. The terms that appear
in the sum represent the virtual transitions between these two modes.
With these two arguments in mind we will make use of the
Born-approximation. In matrix notation,
\begin{equation}
W = (1-W^o)^{-1} W^o \simeq W^o + W^oW^o
\end{equation}

Therefore, in the approximation of small soliton velocity
the terms in (3.28) are small and we can rewrite as a good approximation
\begin{equation}
F[x,y] \simeq \exp\left\{\sum^\infty_{n=1} \Gamma_{nn}[x,y] \;
\overline{n}_n\right\}.
\end{equation}

Observe that if the interaction is turned off $(\Gamma \rightarrow
0)$ or the temperature is zero $(T = 0)$ the functional (3.31) is one,
and, as we would expect the soliton moves as a free particle.

Substituting the Born approximation (3.30) in (3.31) and the latter in (3.12)
we find
\begin{equation}
J = \int^{x}_{x'}{Dx} \int^{y}_{y'}{Dy}\, \exp\left\{\frac{i}{\hbar}
\tilde{S}[x,y] + \frac{1}{\hbar} \tilde{\phi}[x,y]\right\}
\end{equation}
where
\begin{equation}
\tilde{S} = \int^t_0{dt'} \left\{ \frac{M_o}{2}(\dot{x}^2(t') - \dot{y}^2(t'))
+(\dot{x}(t') - \dot{y}(t')) \int^t_0{dt"} \; \Gamma_I(t'-t") (\dot{x}(t") +
\dot{y}(t"))\right\}
\end{equation}
and
\begin{equation}
\tilde{\phi} = \int^t_0{dt'} \; \int^t_0{dt"} \left\{\Gamma_R(t'-t")
 (\dot{x}(t') - \dot{y}(t'))(\dot{x}(t") - \dot{y}(t"))\right\}
\end{equation}
with
\begin{equation}
\Gamma_R(t) = \hbar \theta(t) \sum^\infty_{n,m=1} g^2_{nm}
\overline{n}_n \cos(\Omega_n - \Omega_m)t
\end{equation}
\begin{equation}
\Gamma_I(t) = \hbar \theta(t) \sum^\infty_{n,m=1} g^2_{nm}
\overline{n}_n \sin(\Omega_n - \Omega_m)t.
\end{equation}

Now, if we define the new variables R and r as,
\begin{equation}
R = \frac{x + y}{2}
\end{equation}
\begin{equation}
r = x - y
\end{equation}
the equations of motion for the action in (3.33) read,
\begin{equation}
\ddot{R}(\tau) + 2 \int^t_0{dt'} \gamma(\tau-t') \dot{R}(t') = 0
\end{equation}
\begin{equation}
\ddot{r}(\tau) - 2 \int^t_0{dt'} \gamma(t'-\tau) \dot{r}(t') = 0
\end{equation}
where
\begin{equation}
\gamma(t) = \frac{1}{M_o} \frac{d\Gamma_I}{dt}
\end{equation}
or, using (3.36),
\begin{equation}
\gamma(t) = \frac{\hbar \theta(t)}{M_o} \sum^\infty_{n,m=1}
g^2_{nm} \overline{n}_n(\Omega_n - \Omega_m) \cos(\Omega_n - \Omega_m)t
\end{equation}
is the damping function.

In terms of these newly defined variables, we can easily see that
(3.39) and (3.40) have the same form of the equations previously obtained
in the case of quantum brownian motion \cite{amir}, except for the
fact that they now present memory effects (see ref. \cite{amir} for
details).

Furthermore, in the limit where the time scale of interest is much
greater than the correlation time of the meson variables we can
write \cite{louisell},
\begin{equation}
\gamma(t) = \overline{\gamma}(T) \delta(t)
\end{equation}
where $\overline{\gamma}(T)$ is a damping parameter which is
temperature dependent and $\delta(t)$ is the Dirac delta function.
The form (3.43) is known as the Markovian approximation, because in this case
the memory is purely local and does not depend on the previous motion
of the soliton.

If we use (3.39) and (3.40) with (3.43) and expand the phase of (3.32)
around this classical solution we get the well-known result for the
quantum brownian motion \cite{amir} where the damping parameter
$\gamma$ (temperature independent) is replaced by
$\overline{\gamma}(T)$ and the diffusive part is replaced by (3.34).
As a consequence, the diffusion parameter in momentum space will be
given by,
\begin{equation}
D(t) = \hbar \frac{d^2 \Gamma_R}{dt^2} = -\hbar^2\theta(t)
 \sum^\infty_{n,m=1} g^2_{nm} \overline{n}_n(\Omega_n - \Omega_m)^2
\cos(\Omega_n - \Omega_m)t
\end{equation}
\indent
As discussed before, in the Markovian limit, we can write $D(t)$ in
the Markovian form,
\begin{equation}
D(t) = \overline{D}(T) \delta(t)
\end{equation}
where $\overline{D}(T)$ and $\overline{\gamma}(T)$ obey the classical
fluctuation-dissipation theorem at low temperatures \cite{kubo}.

In what follows we shall define a function $S(\omega, \omega')$ which
will, in analogy to the spectral function $J(\omega)$ of the standard
model \cite{amir}, allows to replace all the summations
over $k$ by integrals over frequencies,
\begin{equation}
S(\omega, \omega') = \sum^\infty_{n,m=1} g^2_{nm} \delta (\omega -
\Omega_n) \delta (\omega' - \Omega_n).
\end{equation}

Notice, however, that unlike $J(\omega)$ in \cite{amir}, this new
function $S(\omega,\omega')$ is related to the scattering of the
environmental excitations between states of frequencies $\omega$
and $\omega'$ (as seeing from the laboratory frame). Moreover, due to
(3.24) it is easy to see that,
\begin{equation}
S(\omega, \omega') = S(\omega', \omega)
\end{equation}
We call $S(\omega,\omega')$ the ``scattering function".

Notice that we can rewrite (3.42) and (3.44) as,
%\[
\begin{equation}
\gamma(t) = \frac{\hbar \theta(t)}{2M_o} \int^\infty_0{d \omega}
\int^\infty_0{d \omega'} S(\omega, \omega') (\omega - \omega')
(n(\omega) - n(\omega'))
%\]
%\begin{equation}
 \cos(\omega - \omega')t
\end{equation}
and
%\[
\begin{equation}
D(t) = -\frac{\hbar^2 \theta(t)}{2} \int^\infty_0{d \omega}
\int^\infty_0{d \omega'}
 S(\omega - \omega')(\omega - \omega')^2
% \]
% \begin{equation}
(n(\omega) + n(\omega')) \cos(\omega - \omega')t.
\end{equation}

Concluding, we have established that the Hamiltonian (2.27) leads to a
brownian dynamics, that is, the soliton moves as a particle in a
viscous environment where its relaxation and diffusion are due to the
scattering of mesons.

\section{Conclusions}

In this article we have showed how a new model for the study of
dissipation in quantum mechanics can be obtained from the study
the solitons in quantum field theory. We showed how the transport
properties of solitons can be calculated in a consistent way
from the microscopic point of view.

Our results show that solitons move as a brownian particle at low energies
due to the scattering of the mesons which are present in the system. These
mesons are the residual excitations created by the presence of the soliton.
We showed that the damping and the diffusion coefficients for the soliton
motion are dependent of the temperature since the mesons must be thermally
activated in order to scatter off the soliton. Therefore, at zero
temperature the soliton moves freely, but its mobility decreases as the
temperature increases (we have showed this explicitly for the case of the
polaron motion, see ref.\cite{prb}). All these results can not be obtained
from the standard model since the origin of the particle and the
environment is distinct is that case.

\acknowledgments

A.~H.~Castro Neto gratefully acknowledges D.~K.~Campbell and E.~Fradkin
for useful comments, R.~Rajaraman for a critical
reading of the manuscript, Conselho Nacional de Desenvolvimento
Cient\'ifico e Tecnol\'ogico, CNPq (Brazil), for a scholarship and
the Department of Physics of the University of Illinois for support
and hospitality. A.O.Caldeira also wishes to acknowledge the partial
support from CNPq.


\newpage

\begin{references}
\bibitem{david}
A.S.Davydov, J.Phys.I {\bf 1}, 1649, (1991).
\bibitem{raja}
R.Rajaraman, {\it Solitons and Instantons}, (North Holland,  Amsterdam, 1982)
and references therein.
\bibitem{amir}
A.O.Caldeira and A.J.Leggett, Physica {\bf 121 A}, 587, (1983).
\bibitem{reif}
F.Reif, {\it Fundamentals of Statistical and Thermal Physics},
(McGraw-Hill, New York, 1965).
\bibitem{prl}
A.H.Castro Neto and A.O.Caldeira, Phys.Rev.Lett., 1960, (1991).
\bibitem{holstein}
T.D.Holstein, Mol.Cryst.Liq.Cryst. {\bf 77}, 235, (1981);
L.A.Turkevich and T.D.Holstein, Phys.Rev.{\bf B35}, 7474, (1987).
\bibitem{prb}
A.H.Castro Neto and A.O.Caldeira, Phys.Rev.{\bf B 46}, 8858, (1992).
\bibitem{feynman}
R.P.Feynman and F.L.Vernon, Annals of Physics {\bf 24}, 118, (1963).
\bibitem{golds}
J.Goldstone and R.Jackiw, Phys.Rev. {\bf D11}, 1486, (1975).
\bibitem{kubo}
R.Kubo, Rep.Prog.Phys. {\bf XXIX}, 253, (1966).
\bibitem{castro neto}
A.H.Castro Neto and A.O.Caldeira, Phys.Rev. {\bf A42}, 6884, (1990).
\bibitem{louisell}
W.H.Louisell, {\it Quantum Statistical Properties of Radiation},
(John Wiley, New York, 1973).
\end{references}
\end{document}